%% file: coldfronts.tex
\documentclass{pazh}

\usepackage{apjfonts}
\usepackage{textcomp}

\def\textdegree{\ensuremath{^\circ}}

\usepackage{pgpic}
\usepackage{multirow}

\def\deg{\ensuremath{^\circ}}

\def\arcmin{\ensuremath{'}}
\def\aprime{a\raise -0.2ex \hbox{$^\prime$}}

\def\muG{\ensuremath{\text{\textmu G}}}
\def\inf{_{\scriptscriptstyle \infty}}
\def\Im{{\rm Im}}
\def\Re{{\rm Re}}

\makeatletter
\def\const  {\mathop{\operator@font const}\nolimits}
\def\asinh   {\mathop{\operator@font asinh}\nolimits}
\makeatother

\begin{document}

\journalinfo{2002}{28}{8}{495}[508]

\title{A Cold Front in A3667: Hydrodynamics and
  Magnetic Field in the Intracluster Medium}

\author{
  A. Vikhlinin\address{1,2},
  M. Markevitch\address{2},
  \addresstext{1}{Institute for Space Research, Moscow, Russia}
  \addresstext{2}{Harvard-Smithsonian Center for Astrophysics}
  }
\shortauthor{VIKHLININ \& MARKEVITCH}
\shorttitle{COLD FRONT IN A3667}
\submitted{\today}

\begin{abstract}

  This conference presentation discusses a \emph{Chandra} observation of the
  cold front in Abell 3667.  We first review our earlier results which
  include a measurement of the front velocity, $M\approx1$, using the ratio
  of exterior and interior gas pressures; observations of the hydrodynamic
  effects expected for a transonic front motion (weak bow shock and gas
  compression near the leading edge of the front); direct observation of the
  suppressed diffusion across the front, and estimate of the magnetic field
  strength near the front from suppression of the Kelvin-Helmholtz
  instabilities.

  The new results include using the 2-dimensional brightness distribution
  inside the cold front (a) to show that the front is stable and (b) to map
  the mass distribution in the gas cloud. This analysis confirms the
  existence of a dark matter subcluster traveling with the front. We also
  fix an algebraic error in our published calculations for the growth rate
  of the KH instability and discuss an additional effect which could
  stabilize the front against the small-scale perturbations. These updates
  only strengthen our conclusions regarding the importance of the magnetic
  fields for the front dynamics.

\end{abstract}

A3667 shows clear signs of the on-going merger of two big subclusters in its
optical, X-ray, and radio images (Sodre et al.\ 1992, Knopp et al.\ 1996,
R\"ottgering et al.\ 1997). The optical image contains two distinct galaxy
concentrations around the giant elliptical galaxies (we call them A and B);
the weak lensing observations reveal the associated mass concentrations
(Joffre et al.\ 2000).  The X-ray emission is elongated and co-aligned with
the galaxy distribution. The most striking feature in the X-ray image is a
strong surface brightness edge to the South-East of galaxy A, located
perpendicularly to the A-B axis. The radio map reveals giant diffuse source
of a 0.5--1~Mpc size, located in the cluster outskirts and oriented
perpendicularly to the A-B axis. All available data point to identification
of galaxies A and B as the centers of the two merging sublclusters. Although
it is expected that sublusters should move supersonically
($v=2000-3000$~km~с$^{-1}$) during the merger, the observed line-of-sight
velocity difference of galaxies A and B is only 120~km~s$^{-1}$ (Katgert et
al.\ 1998). This means that the merger axis is perpendicular to the line of
sight, which significantly simplifies interpretation of the data.

In this conference paper, we first review some earlier results from the
\emph{Chandra} observation of A3667 presented in Vikhlinin et al.\
(2001ab). We also present new results on mapping of the dark mass of the
subcluster traveling with the cold front. We also fix an algebraic error in
the calculations for the growth rate of the KH instability and discuss an
additional effect which could effectively stabilize the front against the
small-scale perturbations.

\section{\emph{Chandra}\, observation of A3667}
\label{sec:a3667:chandra:obs}

The central part of A3667 was observed by \emph{Chandra} in late 1999. The
field of view was centered on the surface brightness edge 8\arcmin{} to the
South-East of galaxy A. \emph{Chandra} angular resolution of $1''$
corresponds to a proper size of 1.46~kpc at the cluster redshift, $z=0.055$
(we assume $H_0 = 50$~km~s$^{-1}$~Mpc$^{-1}$ hereafter). The technical
aspects of the \emph{Chandra} data analysis are presented in Vikhlinin et
al.\ (2001a).

\begin{figure}[t]
  \centerline{
    \LabelSize{\footnotesize}
    \def\fivehundrkpc{{\footnotesize 500 kpc}}
    \XAxisLabelSep{2.5}
    \YAxisLabelSep{0.7}
    \setscale{0.8}
    \input{edge_tex}\hspace*{2em}
    }

  \bigskip

  \centerline{
    \LabelSize{\footnotesize}
    \def\keV{{keV}}
    \XAxisLabelSep{2.5}
    \YAxisLabelSep{0.7}
    \setscale{0.8}
    \input{a3667-tmap_tex}\hspace*{2em}
    }

  \caption{
    \emph{Chandra} X-ray image and temperature map of A3667. Note that the
    surface brightness edge (cold front) is very sharp within the sector
    $2\varphi_{\rm cr}=60\deg$, beyond which it is gradually smeared out.}
  \label{fig:a3667:ch}\label{fig:a3667:tmap}
\end{figure}

\begin{figure}
  {\sffamily\itshape
    \LabelSize{\footnotesize}
    \def\galaxyA{Galaxy A}
    \def\coldfront{cold front}
    \def\kpc{kpc}
    \def\bowshock{possible bow shock}
    \centerline{\setscale{0.7}\input{img_smo_tex}}
    }
  \vskip -10pt
  \caption{}
  \label{fig:a3667:img:smo}
\end{figure}

The \emph{Chandra} X-ray image in the 0.7--4 keV energy band in shown in
Fig.~\ref{fig:a3667:ch}. Figure~\ref{fig:a3667:img:smo} shows its slightly
smoothed version with the main structures marked. The most prominent feature
is a sharp surface brightness edge whose shape is circular with a radius of
$410\pm15\,$kpc. The edge is very sharp --- it is unresolved even with
\emph{Chandra}'s angular resolution.

The temperature map is presented on the bottom panel of
Fig.~\ref{fig:a3667:tmap}. All variations of the temperature in region to
the North-West of the edge, where the surface brightness is high, are
significant. To the South-East of the edge the surface brightness is low and
the temperature uncertainties are quite large. The very hot spots
($T>12$~keV) are statistically insignificant. However, it is very clear that
the gas temperature exterior to the edge is much higher than that
inside. Such structures, observed previously in A2142 (Markevitch et al.\
2001) were termed the ``cold fronts''.

To the South of the cold front, there is a second, weaker surface brightness
discontinuity clearly visible in the smoothed image
(Fig.~\ref{fig:a3667:img:smo}). We discuss below that this structure can be
associated with the bow shock in front of the cold cloud moving at a
slightly supersonic speed. 

\begin{figure*}
  \LabelSize{\footnotesize}\XAxisLabelSep{2.0}
  \figwidth=0.45\linewidth \figheight=\figwidth
  \def\kpc{kpc}
  \def\Sx{$S_x$, erg~s$^{-1}$~cm$^{-2}$/arcmin$^2$}
  \def\keV{keV}

  \noindent
  \input{a3667-prof-br_tex}
  \hfill
  \input{a3667-prof-T_tex}

  \vskip -1ex
  \caption{X-ray surface brightness and temperature profiles across in the $\pm
    15\deg$ sector across the front. The dashed line in the left panel
    corresponds to the \emph{ROSAT} profile without any renormalization. The
  dotted line corresponds to projection of the spherical density discontinuity.}
  \label{fig:a3667:profs}
\end{figure*}
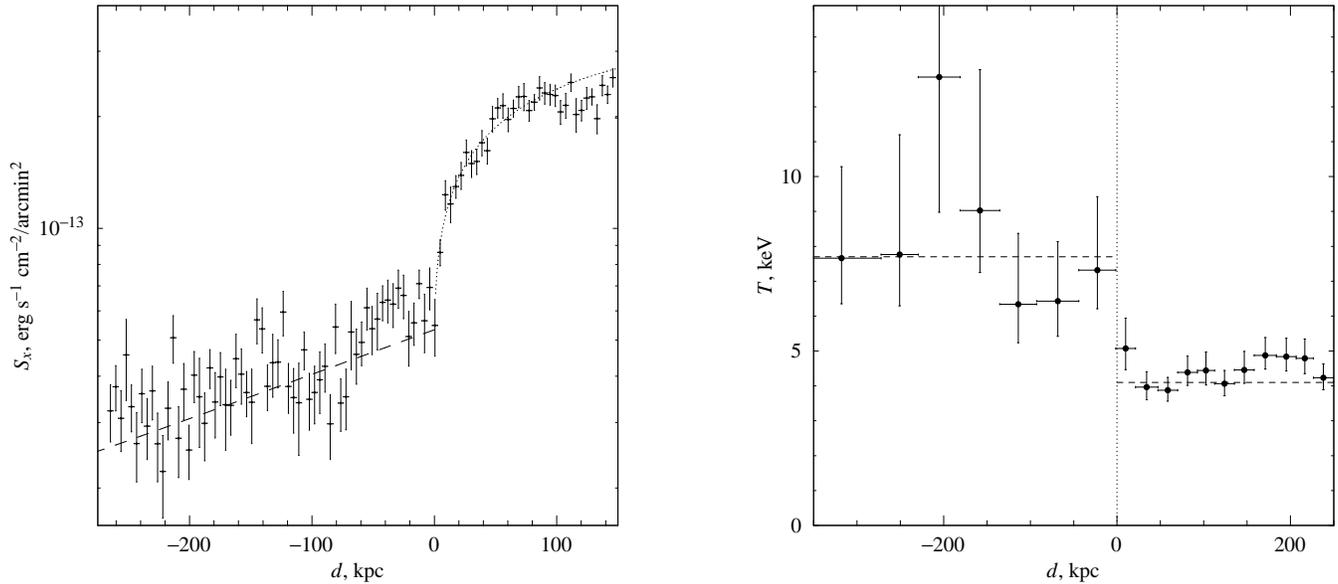

A quantitative study of the cold front properties can be made using the
observed X-ray surface brightness and temperature profiles across the front
(Fig.~\ref{fig:a3667:profs}). The brightness profile shows a sharp
discontinuity. The brightness increases by a factor of two within 7--10~kpc
of the front position and gradually increases by another factor of 2 within
40--80 kpc. Such a shape is typical for the projected spherical density
discontinuity. The deprojection of the surface brightness profile gives the
gas density $n_e=(3.2\pm0.48)\times10^{-3}\,$cm$^{-3}$ interior to the
front, where the error bars conservatively include both measurement and all
systematic uncertainties. Exterior to the front, the gas density is
$n_e=0.82\times10^{-3}\,$cm$^{-3}$.

Across the front, the temperature changes sharply from $\sim 8$ to 4.5 keV
(Fig.~\ref{fig:a3667:profs}). The measured temperatures on both sides are
consistent with constant values within some distance from the front. The
average temperature within 275~kpc exterior to the front is $T_{\rm
out}=7.7\pm0.8$~keV, and within 125~kpc interior to the front it is $T_{\rm
in} = 4.1\pm0.2$~keV.

Therefore, both gas density and temperature undergo a jump across the
surface brightness edge. The corresponding pressure ratio is $p_{\rm
  in}/p_{\rm out} = 2.1\pm0.5$ (Table~\ref{tab:a3667:gas:parameters}), with
a rather conservative errorbar. If the cold front were at rest, there would
be a pressure equilibrium, $p_{\rm in}=p_{\rm out}$. The higher inner
pressure means that the front is moving and therefore the interior gas
``feels'' both the thermal and ram pressure of the exterior gas.

\begin{table}[b]
  \footnotesize
  \def\nodata{\ldots}
  \def\arraystretch{1.2}
  \caption{Gas parameters near the cold front in
    A3667}\label{tab:a3667:gas:parameters} 
  \medskip \centering
  \begin{tabular}{lccc}
    \hline
    \hline
    \multirow{2}{0.7in}{Region}  & $T$ & $n_e$ & $p=Tn_e$\\
            & keV & $10^{-3}\,$cm$^{-3}$ & $10^{-2}\,$keV~cm$^{-3}$\\
    \hline
    Exterior \dotfill & $7.7\pm0.8$ & $0.82\pm0.12$ & $0.63\pm0.11$ \\
    Interior \dotfill & $4.1\pm0.2$ & $3.2\pm0.5$   & $1.32\pm0.21$ \\
    \hline
  \end{tabular}
\end{table}

\section{Hydrodynamics near the cold front}

The observed ram pressure leads to a rather accurate measurement of the
front velocity, $M_1=1.0\pm0.2$, i.e.\ the gas cloud moving at the sound
speed of the hotter gas. The cloud velocity in physical units is
$v=1430\pm290$~km~s$^{-1}$.

Since the cold front moves with a velocity close to the speed of sound, the
compressibility of the gas should be important and therefore two additional
structures can be expected, and indeed observed (Vikhlinin et al.\ 2001a):
a) surface brightness enhancement due to gas compression near the leading
edge of the body, and b) a bow shock if the front velocity is even slightly
supersonic.

\subsection{Is the front shape stable?}
\label{sec:a3667:frond:solid:body}

In the discussion above, we implicitly assumed that the front moves as a
solid body. Can the assumption of the constant front shape be justified?  If
the hydrodynamic instabilities\footnote{In this case the most important type
of instability is Rayleigh-Taylor} are allowed to operate freely, they
significantly disturb (and even destroy) a gas cloud of radius $R$ by the
time it travels a distance of several $R$'s (Jones et al.\ 1996). Since we
observe a very regular shape of the leading edge of the cold front, this
suggests that the instabilities should be somehow suppressed.

Another indication of the constancy of the front shape can be obtained by
considering a dependence of gas pressure on angle from the direction of
motion. All flow lines in the immediate vicinity of the front pass through a
very small region near the leading edge. If the front shape is constant, the
flow is stationary and therefore the gas pressure should follow the
Bernoulli equation --- the pressure decreases as the flow accelerates away
from the stagnation region. The pressure inside the front should correspond
to the exterior pressure, otherwise the shape will change and flow will not
be stationary.  Therefore, the assumption of the constant front shape leads
to an easy and verifiable prediction of the interior pressure distribution
as a function of angle from the direction of motion.

\begin{figure}[tb]
  \figwidth=0.9\linewidth\figheight=\figwidth
  \LabelSize{\footnotesize}
  \XAxisLabelSep{3.0}
  \centerline{\input{azi-p_tex}}
  \vskip -10pt
  \caption{Gas pressure inside the front as a function of angle from the
    leading edge. The line is not the fit, but prediction for the $M=1.05$
    flow.}
  \label{fig:a3667:azi:p}
\end{figure}
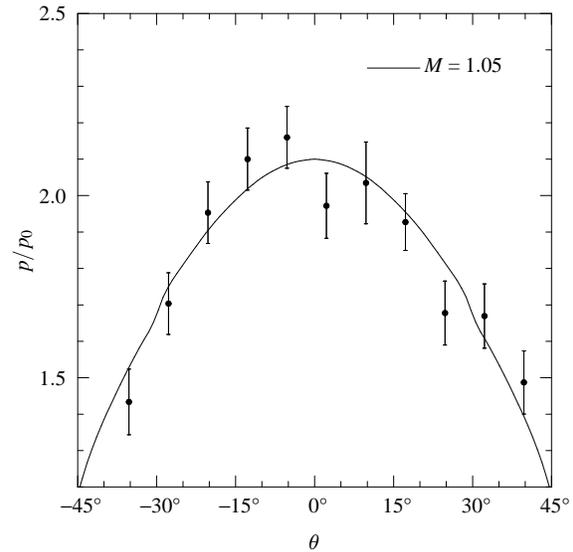

To determine the distribution of gas pressure, we need to measure both gas
density and temperature as a function of angle, $\theta$. We find that
$T=3.8\pm0.40$~keV in the $-10\deg<\theta<10\deg$ sector relative to the
motion direction, $3.8\pm0.55$~keV in the $-30\deg<\theta<-10\deg$ sector,
and $4.7\pm0.70$~keV for $10\deg<\theta<30\deg$.\footnote{The spectra were
collected from a 70~kpc strip just inside the front; angles are front North
to South through East} Within the uncertainties, these values agree with the
mean temperature inside the front $T=4.1\pm0.2$~keV
(Table~\ref{tab:a3667:gas:parameters}), therefore we assume that the
interior gas temperature is constant as a function of angle.

Given the constant temperature, the pressure is simply proportional to the
gas density which is easily derived from deprojection of the surface
brightness profiles across the front in different sectors from the direction
of motion. The resulting angular dependence of the gas pressure in shown in
Fig.~\ref{fig:a3667:azi:p}. The pressure significantly declines --- by a
factor of 1.5 --- at $\theta=30\deg$, where the front starts to widen.
Remarkably, the numerical gasdynamic solution (using the VH-1 code) for the
flow about the solid body at Mach numbers just above 1 predicts just such a
change in pressure (solid line in Fig.~\ref{fig:a3667:azi:p}). Thus, the
angular dependence of the interior gas pressure at least implicitly confirms
that the front moves with a constant shape.

Note that if we can assume \emph{a priori} that the front shape is stable
and the gas temperature inside is uniform, the dependence $p(\theta)$ is a
convenient diagnostics of the front speed. Indeed, $p(\theta)$ is rather
sensitive to the flow speed; for example, the shape of the measured angular
dependence of pressure shown in Fig.~\ref{fig:a3667:azi:p} can be reproduced
only for $0.7<M<1.5$. Under the assumption of constant temperature,
$p(\theta)$ is equivalent to $\rho(\theta)/\rho(0)$, which is easily derived
from the imaging data only.

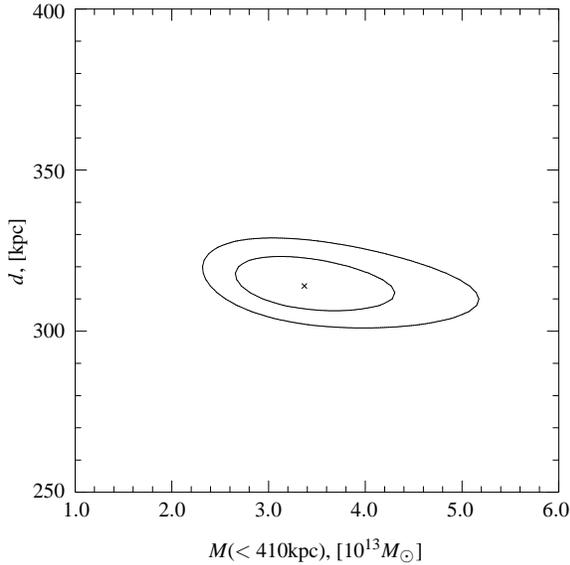
\begin{figure}
  \LabelSize{\footnotesize}
  \XAxisLabelSep{3.0}
  \centerline{\setscale{0.9}\input{a3667halo_tex}}
  \vskip -5pt
  \caption{Confidence intervals(68\% and 95\%) for location of the centroid
    of the King profile $d$ (distance from the center of curvature of the
    cold front), and total mass within $r=410$~kpc.}
  \label{fig:a3667:halo}
\end{figure}

\subsubsection{Dark halo associated with the cold front}
\label{sec:a3667:front:dark:halo}

Since the shape of the cold front remains constant, we can naturally assume
that the interior gas is in hydrostatic equilibrium in the gravitational
field in its reference frame. In this case, the hydrostatic equilibrium
equation, which under assumption of constant temperature reduces to
\begin{equation}\label{eq:a3667:halo:hydrostat}
  \rho = \rho_0 \;\exp(-\mu m_p \varphi / T),
\end{equation}
allows to infer the distribution of mass of the subcluster traveling with
the front. We assume that the subcluster dark matter density profile is
given by the King profile, $\rho_m = \rho_{m0}(1+r^2/r_c^2)^{-3/2}.$ The
density distribution of the cold gas is then given by
equation~(\ref{eq:a3667:halo:hydrostat}) where $\varphi = 4\pi G \rho_0
\left(1-\asinh x/x\right)$, denoting $x=r/r_c$.  Integration of the plasma
emissivity, $\varepsilon\propto\rho^2$, along the line of sight gives the
model distribution of the surface brightness.

The maximum likelihood fit of this model to the observed 2-dimensional
surface brightness distribution within approximately 200~kpc of the front
gives the parameters of the King profile --- the distance, $d$, of the dark
halo centroid from the center curvature of the front, its core-radius $r_c$,
and central density which we parameterize as the total mass within a sphere
of the 410\,kpc radius (the radius of the front). The fit is restricted to
the region near the leading edge of the body due to the following reasons:
1) this region is most interesting from the point of view of the front
dynamics; 2) we can reasonably assume that the gas distribution there is
axially symmetric; 3) this region is small enough for the King profile to
have sufficient freedom for approximating any reasonable distribution of
mass.

\begin{figure}
  \vskip -20pt
  {
    \LabelSize{\footnotesize}
    \sffamily\itshape
    \def\front{{cold front}}
    \def\equipotential{{equipotential surfaces}}
    \def\pdecrease{{$\varphi$ increases, $p$ decreases}}
    
    \centerline{\setscale{0.85}\input{equipot_tex}}
    }
  \caption{}\label{fig:a3667:equipot}
\end{figure}

\begin{figure*}
  \LabelSize{\footnotesize}
  \figwidth=\linewidth\figheight=0.5\figwidth
  \def\sub#1{$_{\text{\sffamily #1}}$}
  \def\vttext{\psframebox*[framesep=0pt]{\sffamily\itshape v \,{\upshape\textgreater} v\sub{\upshape cr}}}
  \centerline{\sffamily\itshape\input{mag_tex}}
  \caption{\emph{(a)} Illustration of the formation of the magnetic
    layer near the front surface. The initially tangled magnetic lines in
    the ambient hot gas (red) are stretched along the surface because of
    tangential motion of the gas. The magnetic lines inside the front are
    stretched because, in the absence of complete magnetic isolation, the
    cool gas experiences stripping. This process can form a narrow layer in
    which the magnetic field is parallel to the front surface. Such a layer
    would stop the transport processes across the front, as well as further
    stripping of the cool gas. \emph{(b)} The interface between the cool and
    hot gas is subject to the Kelvin-Helmholtz instability. The magnetic
    layer can suppress this instability in the region where the tangential
    velocity is smaller that a critical value $v_{\rm cr}$. The velocity
    field shown for illustration corresponds to the flow of incompressible
    fluid around a sphere.}\label{fig:a3667:mag}
\end{figure*}
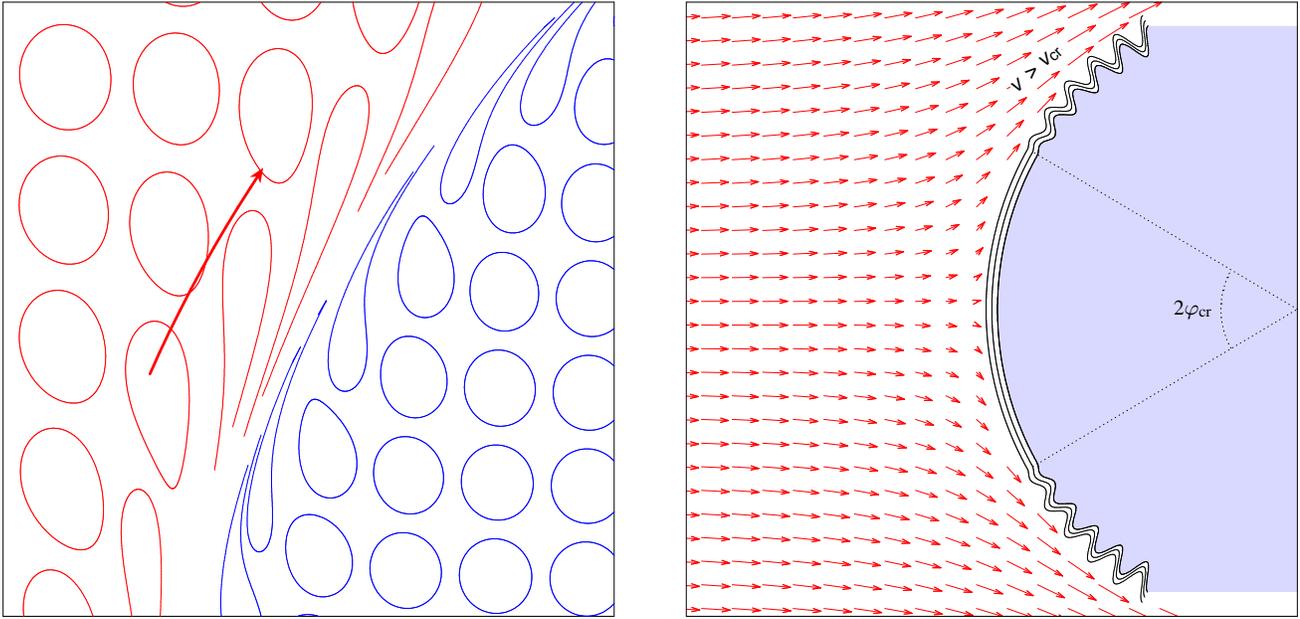

The results of the maximum likelihood fitting are shown in
Fig.~\ref{fig:a3667:halo}. We find that the center of mass is located at
$d=315\pm7$~kpc from the center of curvature of the cold front (i.\,e.\ only
at 95~kpc from the leading edge), and the total mass within 410~kpc is
$(3.2\pm0.8)\times10^{13}\,M_\odot$. The allowed values of the core-radius
are in the range $50\text{~kpc} < r_c < 200\text{~kpc}$. Interestingly, the
derived values of mass and core-radius are very reasonable for the central
region of a 3--4~keV cluster. The total mass within a sphere of the 100~kpc
radius (the distance to the front) is $4.6\times10^{12}\,M_\odot$, with
exceeds the gas mass within the same radius by a factor of 15.

The distance from the subcluster center of mass to the leading edge of the
front is only 25\% of the front radius. This means, roughly speaking, that
the dark matter halo moves in front of the cold gas and pulls it along. Note
that such a configuration is required for the global stability of the front,
as illustrated in Fig.~\ref{fig:a3667:equipot}. Indeed, in the absence of
any gravitational field, the gas pressure inside the front should be
constant, while the exterior pressure should decrease from the leading edge
to the side according to the Bernoulli equation. Therefore, the front shape
could not be stable. The gravitational potential of the dark matter halo
allows a stable non-constant pressure distribution of the interior gas to
emerge. It also provides stability against the global perturbations of the
front shape.  Indeed, let the front bend so that its curvature
increases. The exterior pressure at the outskirts will decrease because the
flow speed increases. At the same time, the interior pressure will increase
since the front will be at a deeper gravitational potential level. The
resulting pressure difference will restore the original front shape.

The presence of the massive dark halo is necessary also for suppression of
the small-scale Rayleigh-Taylor instabilities. Due to ram pressure of the
ambient gas, the cold cloud slows down, which leads to the effective gravity
force directed from the inside of the cold front outward. Since the ambient
gas density is lower, this configuration is unstable. Let us make numerical
estimates. The drag force on the cloud is $F_{\text{d}} = C \times
\rho_{\text{out}} v^2 A/2$, where $\rho_{\text{out}}$ is the density of the
ambient gas, $A$ is the cloud cross-section area, and $C\approx 0.4$ is the
drag coefficient for a cylinder with a rounded head at transonic speed. The
cloud acceleration is then
\begin{equation}
  g_{\text{d}} = \frac{F_{\text{d}}}{M} \approx
  \frac{0.2\,\rho_{\text{out}} \, v^2\, \pi r^2}{4/3\, \pi r^3\,
    \rho_{\text{in}}} = 0.15 \frac{\rho_{\text{out}}}{\rho_{\text{in}}}
  \frac{v^2}{r}
\end{equation}
Substitution of numerical values yields $g_{\text{d}} \approx
8\times10^{-10}\,$cm~s$^{-2}$. At the same time, the gravitational
acceleration near the front due to the subcluster mass is
\begin{equation}
  g = \frac{G M(<100\,\text{kpc})}{(100\,\text{kpc})^2} =
  6.4\times10^{-9}\,\text{cm s$^{-2}$},
\end{equation}
acting in the opposite direction, and therefore the acceleration due to drag
is negligible. Note that in the absence of dark matter, the gravitational
acceleration (due to the gas only) would be a factor of 15 smaller, which is
insufficient to compensate the drag acceleration. Therefore a purely gas
could would be Rayleigh-Taylor unstable, as is indeed seen in numerical
experiments (Jones et al. 1996).

\section{Magnetic structure near the cold front}

Magnetic fields can profoundly affect the properties of the intergalactic
medium (IGM) in clusters. The current measurements of intracluster magnetic
fields are based on Faraday rotation in radio sources seen through the IGM
(e.g., Kim, Kronberg \& Tribble 1991), or on combined radio and hard X-ray
data on cluster radio halos under the assumption that the X-rays are
produced by inverse Compton scattering of the microwave background (e.g.,
Fusco-Femiano et al.\ 2000).  Both these methods indicate the magnetic field
strength on the microgauss level, with considerable uncertainty. The
properties of the cold front in A3667 allow to determine the magnetic field
strength by a completely new method, from its effect on dynamics of the
intracluster gas.

The idea is as follows. We have seen that the cold front is very sharp and
has a smooth shape; therefore, it must be mechanically stable. At the same
time, it can be shown that given the observed velocity of the ambient gas
flow, the sharp front must be rapidly destroyed by the Kelvin-Helmholtz
instability. An examination of the possible mechanisms for the suppression
of this instability points to the magnetic field surface tension as the most
probable candidate.

\label{a3667:mag:layer:creation}
The required magnetic field configuration can form through a scenario
schematically shown in Fig.~\ref{fig:a3667:mag}. An initially random
magnetic field is frozen in the intracluster gas. Because of the tangential
plasma motions around the front, the magnetic loops are stretched along the
surface. The magnetic reconnection can then produce a layer in which the
field lines are parallel to the interface between the cold and hot gases and
to the flow. This configuration of the magnetic field also provides surface
tension that can suppress the KH instability.

\label{kh:mag:idea}
The flow speed increases as a function of distance from the cold front. If
the magnetic field is not too strong, sooner or later it will be unable to
stabilize the KH perturbations, and at this point the instabilities will
start to grow; the location of this point is marked as $\varphi_{\rm cr}$ is
Fig.~\ref{fig:a3667:mag}. We expect that within the sector $\pm\varphi_{\rm
cr}$ is sharp, while beyond it, the cloud boundary is smeared by
turbulence. One readily notes the analogy with the X-ray image of A3667,
where the front is sharp within $\varphi\approx30\deg$ and quickly smears
beyond this radius (Fig.~\ref{fig:a3667:ch}). Assuming that at
$\varphi=30\deg$, the magnetic field tension is just enough to provide
mechanical stability, we can determine the magnetic field strength.

Let us consider in detail the line of argument. For simplicity, we assume
that the front velocity is $M = 1$ exactly.

\subsection{Hydrodynamic KH instability of the cold front}

The interface between two fluids in most cases is subject to two types of
hydrodynamic instability: (i) Rayleigh-Taylor instability, which arises if
the effective gravity is directed from the heavier fluid into the lighter
one; and (ii) Kelvin-Helmholtz instability, which arises from tangential
motion of the fluids. As was discussed above, the dark matter halo moving
with the front suppresses the Rayleigh-Taylor instability. Therefore, we
consider below only the KH instability.

A numerical solution of the $M=1$ flow about the cold front gives the
following angular dependence of the velocity of the ambient gas near the
front:
\begin{equation}\label{eq:rizzi:m}
  M(\varphi) \simeq1.1\sin\varphi.
\end{equation}
At $\varphi=30\deg$ the Mach number reaches $M=0.55$. Clearly, this is a
high velocity and therefore one expects a fast growth of the KH
perturbations leading to formation of a turbulent layer. Below, we will use
the following widely known properties of the KH instability: 1) The
fastest-growing perturbations are parallel to the flow; 2) The
short-wavelength modes grow faster than the long-wavelength modes; 3) The
perturbations grow exponentially, and therefore a nonlinear stage is reached
rapidly. When the nonlinear stage is reached, the amplitude is approximately
equal to the wavelength of the initially linear perturbation; subsequently,
the perturbations with a given wavelength merge into large vortices. Thus,
if one can show that the perturbations with wavelength $\lambda$ are
unstable, the turbulent layer of width at least $\lambda$ is expected.

\subsubsection{Dispersion equation}

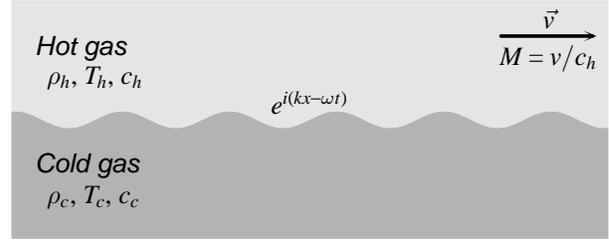
\begin{figure}
  {
    \def\Hot{\sffamily\itshape Hot gas}
    \def\Cold{\sffamily\itshape Cold gas}
    \def\Hotgas{$\rho_h$, $T_h$, $c_h$}
    \def\Coldgas{$\rho_c$, $T_c$, $c_c$}
    \setscale{0.75}
    \centerline{\input{kh-disp-sketch_tex.tex}}
    }
  \caption{Hot gas flows with a velocity $\vec{v}$ relative to 
    the colder, denser gas. The interface is subject to the planar wave
    perturbations, $\exp(i(kx-\omega t))$, with the wave vector parallel to
    the flow. For the given wave vector, $\omega$ can be found from the
    dispersion equations: (\ref{eq:dispers}) for pure hydrodynamic and
    (\ref{eq:dispers:mag}) in MHD.}\label{fig:kh:disp:sketch}
\end{figure}

The stability analysis is conveniently performed via consideration of the
small perturbation of the interface in the form of the planar waves and
solving the resulting dispersion equation. In our case the dispersion
equation can be written as (Miles 1958):
\begin{equation}\label{eq:dispers}
  -\frac{1}{\omega^2}-\frac{c_h^2/c_c^2}{(\omega-Mc_hk)^2}+
  \frac{1}{k^2c_c^2}=0,
\end{equation}
where $M$ is the Mach number of the hot gas, and $c_c$ and $c_h$ sound
speeds in the cold and hot gases, respectively (see
Fig.~\ref{fig:kh:disp:sketch}). 

\subsubsection{Application to A3667}

For all relevant Mach numbers, equation~(\ref{eq:dispers}) has two complex
solutions for $\omega$, one of them corresponding to the growing mode. The
instability growth time is $\tau=(\Im\,\omega)^{-1}$. Although perturbations
with any wavelength are formally unstable, only for some of them the growth
time is sufficiently short compared to the lifetime of the front. The
relevant time scale with which $\tau$ should be compared is the cluster core
passage time, $t_{\rm cross}=L/M\inf c\inf$, where $L$ is the cluster
size. The value of $\exp(t_{\rm cross}/\tau)$ is the perturbation growth
factor over the core passage time; if $t_{\rm cross}/\tau\gg1$ (we will use
a conservative fiducial threshold of 10), the perturbation is effectively
unstable because it has enough time to become non-linear.

For the observed gas parameters and $M\inf=1$ the solution of
(\ref{eq:dispers}) is approximately given by $t_{\rm
cross}/\tau\simeq12\,M\, L/\lambda$, where $\lambda=2\pi/k$ is the
perturbation wavelength. Using the distribution of the Mach number near the
front (eq.~\ref{eq:rizzi:m}), we have\footnote{We note that the analogous
equation (4) in Vikhlinin et al.\ 2001b contained a factor of $(2\pi)^2$
smaller numerical factor due to an algebraic error.}
\label{sec:a3667:hydro:inst:timescales}
\begin{equation}\label{eq:tcross/tau}
  \frac{t_{\rm cross}}{\tau} \simeq 3.3\, \frac{L}{\lambda}\, \sin\varphi.
\end{equation}
This equation, applied for a typical cluster size of $L=1$~Mpc, implies that
perturbations with $\lambda=10$~kpc are unstable ($t_{\rm cross}/\tau=10$)
at $\varphi>1.7\deg$, that is, essentially everywhere; at $\varphi=10\deg$,
the modes with $\lambda<57$~kpc are unstable, and at $\varphi=30\deg$, modes
with $\lambda<150$~kpc.

There is another effect which limits the lifetime of perturbations. The
linear perturbations have non-zero group velocity $v_{\text{dr}} =
d\Re\,\omega/dk$, and therefore drift along the flow as they grow. In
principle, in some cases the perturbation can move to the side of the front
before it grows non-linear. A detailed analysis of solutions of
eq.~(\ref{eq:dispers}) shows that $\omega$ is always proportional to $k$ and
therefore the group velocity equals the phase velocity, $v_{\text{dr}} =
\Re(\omega)/k$. Furthermore, in the entire range of the flow velocities near
the cold front in A3667, the imaginary part of $\omega$ is almost
proportional to its real part:
\begin{equation}
  \frac{\Re\,\omega}{\Im\,\omega} \simeq 0.70.
\end{equation}
This equation implies a simple relation between the perturbation growth
factor and its displacement, $\Delta\varphi$:
\begin{equation}
  \begin{split}
    &\text{Growth factor} = \\
    &\exp\left(\int \Im\,\omega\,dt\right) \simeq
    \exp\left(1.43 \int \Re\,\omega\,dt\right)  = \\
    & = \exp\left(1.43\, k \int \frac{\Re\,\omega}{k}\,dt\right) = 
    \exp\left(1.43\, k \int dl\right) = \\
    &\qquad
    \exp\left(4.7\,\frac{R}{\lambda}\,\frac{\Delta\varphi}{30\deg}\right), 
  \end{split}
\end{equation}
where $R$ is the radius of the front. Assuming, as before, that reaching a
growth factor of $e^{10}$ is required for a perturbation to become
non-linear, we obtain the following condition for effective instability for
$R=410$~kpc:
\begin{equation}\label{eq:drift:instability:cond}
  \lambda < 190\,\text{kpc}\;\frac{\Delta\varphi}{30\deg}.
\end{equation}
The longer wavelength perturbations drift to the side of the front before
they grow non-linear. Perturbations with the interesting $\lambda \sim
30$~kpc should grow strongly non-linear before they leave the observed
smooth region of the front. 

To summarize, the perturbations with $\lambda\lesssim30$~kpc are KH-unstable
over most of the front surface, and therefore a formation of the turbulent
layer of at least this width is expected. Since the observed front width is
$<5$~kpc (Vikhlinin et al.\ 2001a), the KH instability must be suppressed.

The commonly known mechanisms for suppression of the KH instability are a)
gravity and b) surface tension at the interface. Let us consider them in
application to A3667.

\subsection{The role of gravity in suppressing the KH instability}

As we showed above, there is a massive subcluster traveling with the front
and its gravity has some stabilizing effect on the front. The gravitational
force suppresses the KH instability if it acts in the direction from the
lighter fluid into the heavier one, and if the following stability condition
is satisfied:
\begin{equation}\label{eq:kh:stabilization:gravity}
  \frac{g}{k} > v^2\,\frac{{{\rho }_c}\,{{\rho }_h}}
  {\rho_c^2 - \rho_h^2}.
\end{equation}
It is clear from this equation that only the long wavelength modes are
suppressed by gravity, while the shorter wavelength perturbations can grow
almost freely. Noting that $\rho_c\gg \rho_h$, equation
(\ref{eq:kh:stabilization:gravity}) can be rewritten as
\begin{equation}
  \frac{g}{k} \gtrsim v^2 \frac{\rho_h}{\rho_c} = v^2
  \frac{T_c}{T_h} = v^2 \frac{T_c}{c_s^2\,\mu m_p/\gamma} = \frac{T_c}{\mu
    m_p/\gamma} M^2.
\end{equation}
The gravitational acceleration in the front rest frame was derived
above. Using its numerical value, we obtain the following stability condition:
\begin{equation}
  \frac{\lambda}{\text{1\,kpc}} > 3500\, M^2 = 4200\, \sin^2\varphi 
\end{equation}
This implies that the perturbations with $\lambda=10\,$kpc are stabilized by
gravity only within the sector $\varphi<3\deg$; at $\varphi=5\deg$ stability
is achieved only for $\lambda>32$~kpc, and at $\varphi=30\deg$ --- for
$\lambda>1050$~kpc.

Therefore, gravity cannot suppress a formation of a turbulent layer with a
width of 10--20~kpc. This leaves the surface tension of the magnetic field
as the most likely mechanism for suppression of the KH instability.

\subsection{Suppression of the KH instability by magnetic field}

The KH instability can be suppressed by the surface tension of a magnetic
field, if the latter is parallel to the interface and to the direction of
the flow. We have discussed above that even if the magnetic field is
initially highly tangled, the tangential gas motion near the front can
create the required magnetic configuration (see.\ Fig.~\ref{fig:a3667:mag}).

The analysis of KH instability in MHD is greatly simplified if the plasma
can be considered incompressible. Fortunately, this can be done in our case
because the gas exterior to the front flows with a relatively low local Mach
number $M\lesssim0.5$, and the growing modes of the KH instability in the
interior cool gas generally have low phase speed. The incompressibility
assumption can be also justified by the fact that the growth time of the
hydrodynamic instability given by eq.~(\ref{eq:dispers}) is very close to
that in the incompressible limit, i.\,e.\ for $M c_h\rightarrow v$,
$c_c/c_h^2\rightarrow \rho_h/\rho_c$, $c_c \rightarrow \infty$.

The dispersion equation for small perturbations in a perfectly conducting,
incompressible plasma can be written as (Syrovatskij 1953)
\begin{equation}\label{eq:dispers:mag}
  \rho_h (\omega-kv)^2 + \rho_c\omega^2 = k^2\left(\frac{B_h^2}{4\pi} +
  \frac{B_c^2}{4\pi}\right)
\end{equation}
where $B_h$ and $B_c$ are the magnetic field strengths in the hot and cold
gas, respectively (we assume that the field lines are elongated with the
flow), and $\rho_h$ and $\rho_c$ are the gas densities. The solutions of
this equation are real (the interface is stable) if
\begin{equation}\label{eq:kh:mag}
  B_h^2+B_c^2 > 4\pi\, \frac{\rho_h\,\rho_c}{\rho_h+\rho_c}\, v^2.
\end{equation}

\subsubsection{Application to A3667}

If the magnetic pressure is small compared to the thermal pressure, $p$ (as
it turns out to be the case), or is the same fraction of $p$ on both sides
of the interface, the stability condition~(\ref{eq:kh:mag}) can be rewritten
in terms of the thermal pressure and temperature of the gas:
\begin{equation}\label{eq:kh:mag:P:condition}
  \frac{B_h^2}{8\pi}+\frac{B_c^2}{8\pi}>\frac{1}{2}\,\frac{\gamma
    M^2}{1+T_c/T_h}  p_{\text{gas}}.
\end{equation}
Therefore, for the observed temperatures and flow velocities the stability
of the cold front within the sector $\varphi<30\deg$, where $M\le0.55$,
requires that $(B_h^2+B_c^2)/8\pi>0.17p$. If smearing of the front beyond
this sector is interpreted as the onset of the instability, this becomes an
estimate of the total magnetic pressure:
\begin{equation}\label{eq:pmag/pgas}
  p_{\text{mag},h} + p_{\text{mag},c} = 0.17 p_{\text{gas}}
\end{equation}


The formal statistical uncertainty in the derived value of the magnetic
field strength is mostly due to uncertainty in the angle, $\varphi_{\rm
cr}$, where the interface becomes unstable, and hence in the local Mach
number of the hot gas. For $\varphi_{\rm cr}$ in the range $30\deg\pm10\deg$
(which appears to be a conservative interval, see Fig.~\ref{fig:a3667:ch}),
we find from eq.~(\ref{eq:kh:mag}) and (\ref{eq:rizzi:m}) that the magnetic
field strengths are in the interval
\begin{equation}\label{eq:pmag:contsraint}
 0.09p<\frac{B_h^2+B_c^2}{8\pi}<0.23p.
\end{equation}
The maximum of the magnetic field strengths in the cold and hot gases,
$B=\max(B_h,B_c)$, is in the interval $(4\pi p_{\rm mag})^{1/2}<B<(8\pi
p_{\rm mag})^{1/2}$, where $p_{\rm mag} = (B_h^2+B_c^2)/8\pi$ is constrained
by (\ref{eq:pmag:contsraint}). Using the value of gas pressure inside the
dense cloud, we find the corresponding uncertainty interval
$7\,\mbox{\muG}<B<16\,\mbox{\muG}$.

Thus, the front sharpness near the axis and its gradual smearing at large
angles are most likely explained by the existence of a layer with a $\sim
10$~\muG\ magnetic field parallel to the
front. Equation~(\ref{eq:pmag/pgas}) shows that the magnetic pressure in the
layer is a small fraction of the gas pressure. Everywhere else it should be
even smaller, because the magnetic field in the layer is amplified by
stretching of the field lines. This conclusion is robust: e.g., if the
magnetic pressure were of order of the thermal pressure,
eq.~(\ref{eq:kh:mag:P:condition}) implies that the front would be stable for
Mach numbers up to $M\sim1.9$, i.e.\ over the entire surface of the cool gas
cloud.

Faraday rotation measurements of the magnetic field outside the cluster
cooling flows provide the values of $B$ from $\sim 1$~\muG\ (Kim et al.\
1991) to $\sim 10$~\muG\ (Clarke et al.\ 2000), with considerable
uncertainty of the absolute field strength due to unknown degree of
entanglement. Near the cold front, the field is straightened and amplified,
so our estimate likely is an upper limit on the absolute field strength in
the cluster core, and thus is consistent with the Faraday rotation
measurements in the bulk of IGM.

Note that to suppress the instability, the magnetic field does not have to
be completely ordered. It is sufficient to arrange for uniformity on the
scales of order 100~kpc. The perturbations with $\lambda<10$~kpc will then
stabilized, while the growth time for perturbations with╩$\lambda>100$~kpc
is longer than the front lifetime anyway.

The stretching of magnetic field lines by tangential plasma motions is not
the only mechanism for creation of a layer with a highly ordered magnetic
field. For example, Frank et al. (1996) and Jones et al. (1997) showed that
even a weak seed magnetic field which is incapable of suppressing the KH
instability, is significantly amplified in the nonlinear vortices, and after
active field line reconnections, a thin layer is formed. In this layer, the
field lines are aligned with the flow and the field strength increases so
that interface becomes stable. Therefore, the formation of a magnetic layer
considered above seems natural; once formed, the layer acts as ``magnetic
isolation'' which stops the transport processes across the front, and also
as ``magnetic lubrication'', which provides for nonviscous, laminar flow
around the cold front.

In conclusion, we note that the structures similar to the cold fronts in
A3667 and A2142 have now been detected in many other clusters; this is a
rather common phenomenon (see Markevitch et al.\ 2002 for a review).

\bigskip

We thank P.~Mazzotta for pointing out an algebraic error in derivation of
equation (6).

\end{document}

%% file: edge_tex.tex
\setlength{\figwidthdef}{ 3.5000000000in}%
\setlength{\figheightdef}{ 3.5000000000in}%
\setuplengths%
 \begin{pspicture}(3600,3600)%
 \rput(1800,1800){\includegraphics[width=\PGscale\psfigwidth,height=\PGscale\psfigheight]{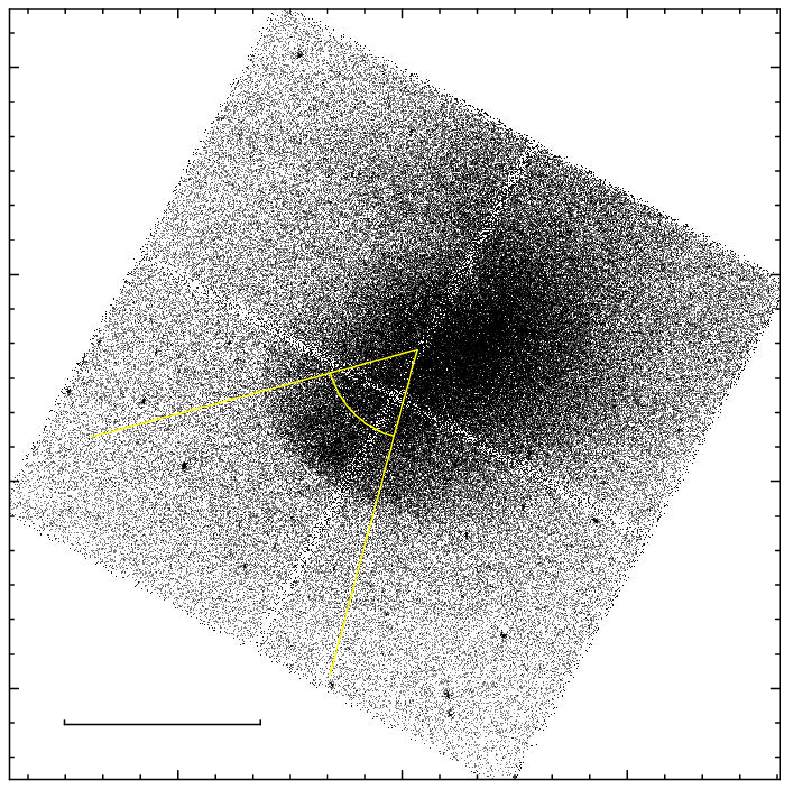}}%
\putlabel{b }{   0.0}{1048}{ 664}{\upshape 500 kpc}%
\def\xticklabeldir{90}%
\setlength{\xtickheight}{0pt}%
\def\xticklabeldir{270}%
\xticklabel{2961}{ 420}{$303.0$\rlap{\textdegree}}%
\xticklabel{2036}{ 420}{$303.2$\rlap{\textdegree}}%
\xticklabel{1112}{ 420}{$303.4$\rlap{\textdegree}}%
\def\yticklabeldir{0}%
\setlength{\ytickwidth}{0pt}%
\def\yticklabeldir{180}%
\yticklabel{ 420}{ 794}{$-57.0$\rlap{\textdegree}}%
\yticklabel{ 420}{1646}{$-56.9$\rlap{\textdegree}}%
\yticklabel{ 420}{2497}{$-56.8$\rlap{\textdegree}}%
\yticklabel{ 420}{3349}{$-56.7$\rlap{\textdegree}}%
\def\xlabeldir{270}%
\PutXAxisLabel{2005}{ 420}{$\alpha$}%
\def\ylabeldir{180}%
\PutYAxisLabel{ 420}{2005}{$\delta$}%
\newrgbcolor{pgcolor}{
     1.00000    1.00000  0.}%
\putlabel{  }{   0.0}{1767}{1864}{\larger $2\varphi_{\rm cr}$}%
\newrgbcolor{pgcolor}{
   0.  0.  0.}%
 \end{pspicture}%

%% file: a3667-tmap_tex.tex
\setlength{\figwidthdef}{ 3.5000000000in}%
\setlength{\figheightdef}{ 3.5000000000in}%
\setuplengths
 \begin{pspicture}(3600,3600)%
 \rput(1800,1800){\includegraphics[width=\PGscale\psfigwidth,height=\PGscale\psfigheight]{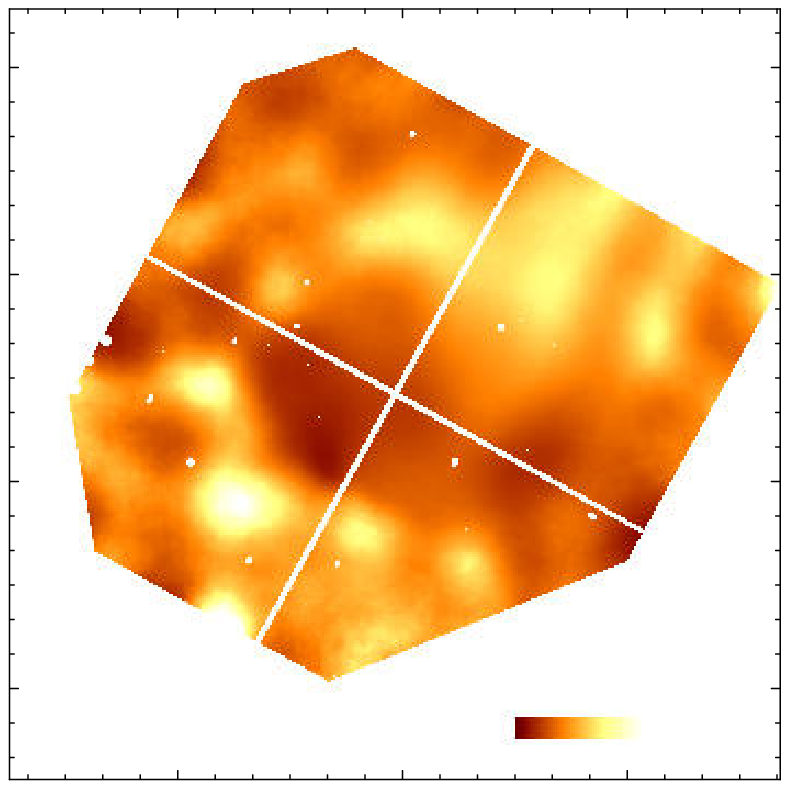}}%
\putlabel{b }{   0.0}{2506}{ 699}{\upshape 3}%
\putlabel{b }{   0.0}{3048}{ 699}{\upshape 15 \rlap{\keV}}%
\putlabel{b }{   0.0}{2777}{ 699}{\upshape 9}%
\def\xticklabeldir{90}%
\setlength{\xtickheight}{0pt}%
\def\xticklabeldir{270}%
\xticklabel{2961}{ 420}{$303.0$\rlap{\textdegree}}%
\xticklabel{2036}{ 420}{$303.2$\rlap{\textdegree}}%
\xticklabel{1112}{ 420}{$303.4$\rlap{\textdegree}}%
\def\yticklabeldir{0}%
\setlength{\ytickwidth}{0pt}%
\def\yticklabeldir{180}%
\yticklabel{ 420}{ 794}{$-57.0$\rlap{\textdegree}}%
\yticklabel{ 420}{1646}{$-56.9$\rlap{\textdegree}}%
\yticklabel{ 420}{2497}{$-56.8$\rlap{\textdegree}}%
\yticklabel{ 420}{3349}{$-56.7$\rlap{\textdegree}}%
\def\xlabeldir{270}%
\PutXAxisLabel{2005}{ 420}{$\alpha$}%
\def\ylabeldir{180}%
\PutYAxisLabel{ 420}{2005}{$\delta$}%
 \end{pspicture}%

%% file: img_smo_tex.tex
\setlength{\xunitlen}{0.0009375000in}%
\setlength{\yunitlen}{0.0009375000in}%
\setlength{\runitlen}{0.0009375000in}%
\psset{xunit=\PGscale\xunitlen}%
\psset{yunit=\PGscale\yunitlen}%
\psset{runit=\PGscale\runitlen}%
 \begin{pspicture}(3600,3600)%
 \rput(1800,1800){\includegraphics[scale=\PGscale]{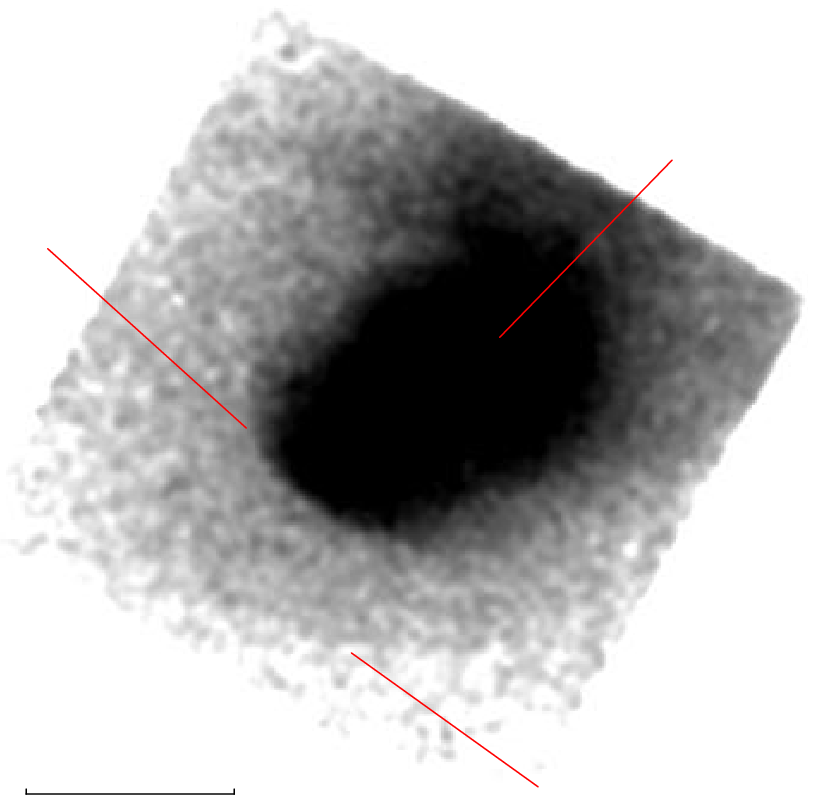}}%
\newrgbcolor{pgcolor}{
     1.00000  0.  0.}%
\newrgbcolor{pgcolor}{
   0.  0.  0.}%
\putlabel{b }{   0.0}{ 355}{2589}{\coldfront}%
\newrgbcolor{pgcolor}{
     1.00000  0.  0.}%
\newrgbcolor{pgcolor}{
   0.  0.  0.}%
\putlabel{b }{   0.0}{2957}{  89}{\bowshock}%
\newrgbcolor{pgcolor}{
     1.00000  0.  0.}%
\newrgbcolor{pgcolor}{
   0.  0.  0.}%
\putlabel{b }{   0.0}{3018}{2936}{\galaxyA}%
\putlabel{b }{   0.0}{ 707}{ 253}{500 \kpc}%
\newrgbcolor{pgcolor}{
   0.    1.00000  0.}%
 \end{pspicture}%

%% file: a3667-prof-br_tex.tex
\setlength{\figwidthdef}{ 3.3750000000in}%
\setlength{\figheightdef}{ 3.3750000000in}%
\setuplengths
 \begin{pspicture}(3600,3600)%
 \rput(1800,1800){\includegraphics[width=\PGscale\psfigwidth,height=\PGscale\psfigheight]{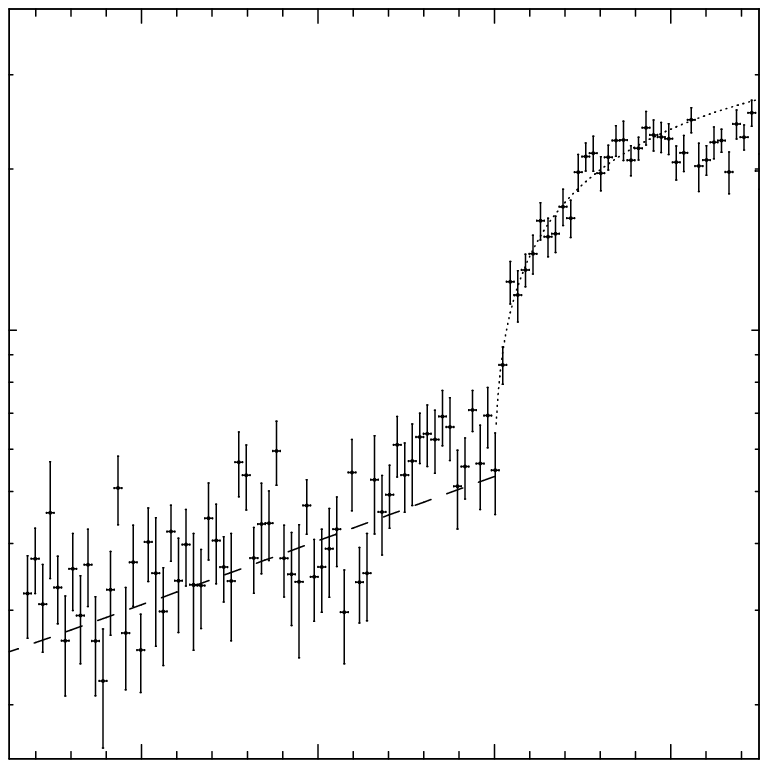}}%
\def\yticklabeldir{0}%
\setlength{\ytickwidth}{0pt}%
\def\yticklabeldir{180}%
\yticklabel{ 400}{2228}{$10\smash{^{-13}}$}%
\def\xticklabeldir{90}%
\setlength{\xtickheight}{0pt}%
\def\xticklabeldir{270}%
\xticklabel{ 965}{ 400}{$-200$}%
\xticklabel{1717}{ 400}{$-100$}%
\xticklabel{2470}{ 400}{$0$}%
\xticklabel{3223}{ 400}{$100$}%
\def\ylabeldir{180}%
\PutYAxisLabel{ 400}{2000}{\Sx}%
\def\xlabeldir{270}%
\PutXAxisLabel{2000}{ 400}{$d$, \kpc}%
 \end{pspicture}%

%% file: a3667-prof-T_tex.tex
\setlength{\figwidthdef}{ 3.3750000000in}%
\setlength{\figheightdef}{ 3.3750000000in}%
\setuplengths
 \begin{pspicture}(3600,3600)%
 \rput(1800,1800){\includegraphics[width=\PGscale\psfigwidth,height=\PGscale\psfigheight]{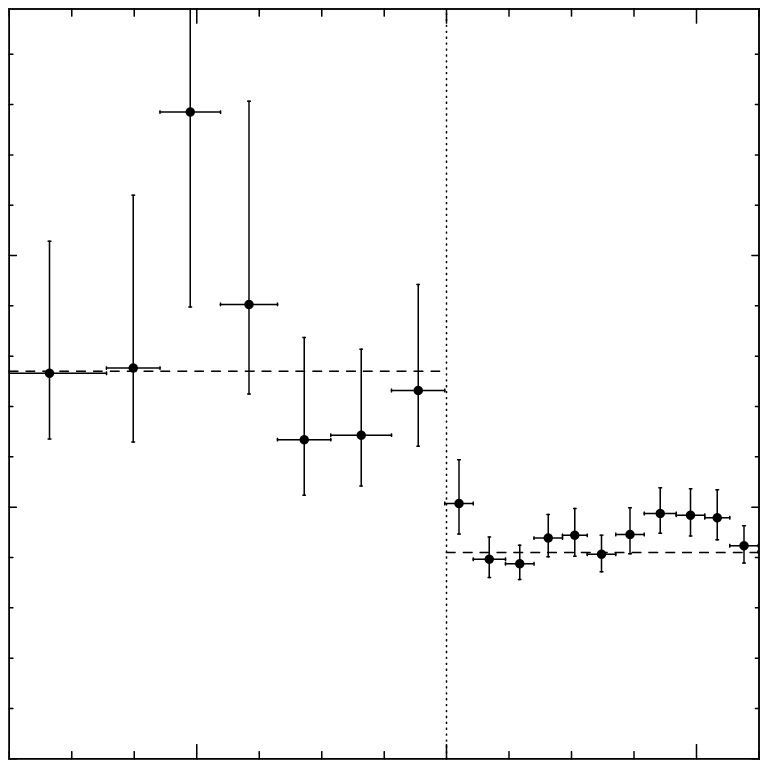}}%
\def\yticklabeldir{0}%
\setlength{\ytickwidth}{0pt}%
\def\yticklabeldir{180}%
\yticklabel{ 400}{ 400}{$0$}%
\yticklabel{ 400}{1473}{$5$}%
\yticklabel{ 400}{2547}{$10$}%
\def\xticklabeldir{90}%
\setlength{\xtickheight}{0pt}%
\def\xticklabeldir{270}%
\xticklabel{1200}{ 400}{$-200$}%
\xticklabel{2266}{ 400}{$0$}%
\xticklabel{3332}{ 400}{$200$}%
\def\ylabeldir{180}%
\PutYAxisLabel{ 400}{2000}{$T$, \keV}%
\def\xlabeldir{270}%
\PutXAxisLabel{2000}{ 400}{$d$, \kpc}%
 \end{pspicture}%

%% file: azi-p_tex.tex
\setlength{\figwidthdef}{ 3.3750000000in}%
\setlength{\figheightdef}{ 3.3750000000in}%
\setuplengths%
 \begin{pspicture}(3600,3600)%
 \rput(1800,1800){\includegraphics[width=\PGscale\psfigwidth,height=\PGscale\psfigheight]{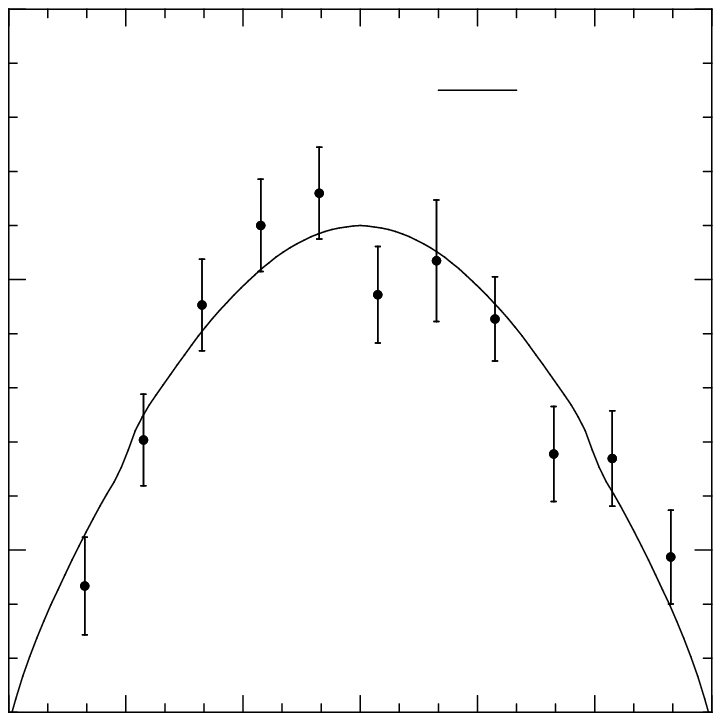}}%
\def\xticklabeldir{90}%
\setlength{\xtickheight}{0pt}%
\def\xticklabeldir{270}%
\xticklabel{ 500}{ 500}{$-45$\rlap{\textdegree}}%
\xticklabel{1000}{ 500}{$-30$\rlap{\textdegree}}%
\xticklabel{1500}{ 500}{$-15$\rlap{\textdegree}}%
\xticklabel{2000}{ 500}{$0$\rlap{\textdegree}}%
\xticklabel{2500}{ 500}{$15$\rlap{\textdegree}}%
\xticklabel{3000}{ 500}{$30$\rlap{\textdegree}}%
\xticklabel{3500}{ 500}{$45$\rlap{\textdegree}}%
\def\yticklabeldir{0}%
\setlength{\ytickwidth}{0pt}%
\def\yticklabeldir{180}%
\yticklabel{ 500}{1192}{$1.5$}%
\yticklabel{ 500}{2346}{$2.0$}%
\yticklabel{ 500}{3500}{$2.5$}%
\def\xlabeldir{270}%
\PutXAxisLabel{2000}{ 500}{$\theta$}%
\def\ylabeldir{180}%
\PutYAxisLabel{ 500}{2000}{$p/p_0$}%
\putlabel{l }{   0.0}{2700}{3176}{$M=1.05$}%
\newrgbcolor{pgcolor}{
   0.  0.  0.}%
 \end{pspicture}%

%% file: a3667halo_tex.tex
\setlength{\xunitlen}{0.0009375000in}%
\setlength{\yunitlen}{0.0009375000in}%
\setlength{\runitlen}{0.0009375000in}%
\psset{xunit=\PGscale\xunitlen}%
\psset{yunit=\PGscale\yunitlen}%
\psset{runit=\PGscale\runitlen}%
 \begin{pspicture}(3600,3600)%
 \rput(1800,1800){\includegraphics[scale=\PGscale]{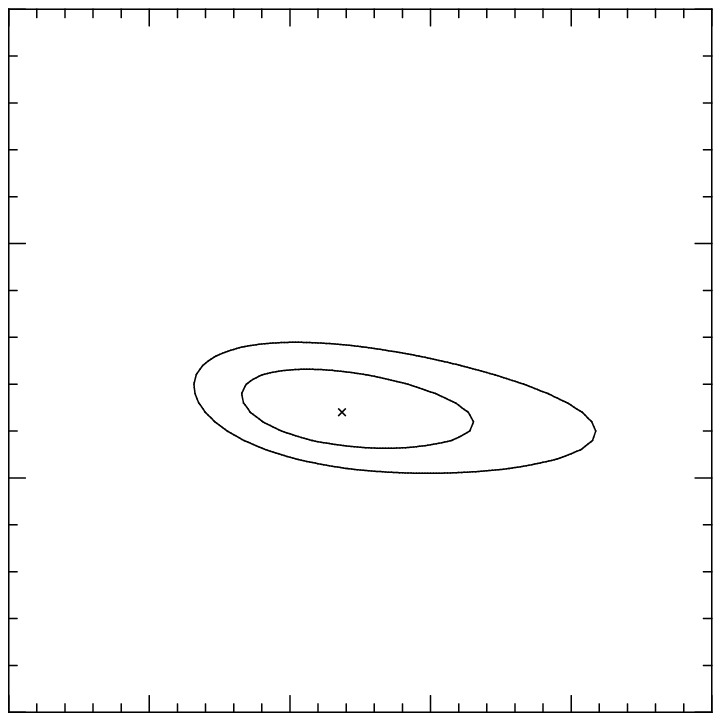}}%
\def\xticklabeldir{90}%
\setlength{\xtickheight}{0pt}%
\def\xticklabeldir{270}%
\xticklabel{ 500}{ 500}{$1.0$}%
\xticklabel{1100}{ 500}{$2.0$}%
\xticklabel{1700}{ 500}{$3.0$}%
\xticklabel{2300}{ 500}{$4.0$}%
\xticklabel{2900}{ 500}{$5.0$}%
\xticklabel{3500}{ 500}{$6.0$}%
\def\yticklabeldir{0}%
\setlength{\ytickwidth}{0pt}%
\def\yticklabeldir{180}%
\yticklabel{ 500}{ 500}{$250$}%
\yticklabel{ 500}{1500}{$300$}%
\yticklabel{ 500}{2500}{$350$}%
\yticklabel{ 500}{3500}{$400$}%
\def\xlabeldir{270}%
\PutXAxisLabel{2000}{ 500}{$M (<410 \text{kpc})$, [$10^{13} M_\odot$]}%
\def\ylabeldir{180}%
\PutYAxisLabel{ 500}{2000}{$d$, [kpc]}%
 \end{pspicture}%

%% file: equipot_tex.tex
\setlength{\xunitlen}{0.0009722222in}%
\setlength{\yunitlen}{0.0009722222in}%
\setlength{\runitlen}{0.0009722222in}%
\psset{xunit=\PGscale\xunitlen}%
\psset{yunit=\PGscale\yunitlen}%
\psset{runit=\PGscale\runitlen}%
 \begin{pspicture}(3600,3600)%
 \rput(1800,1800){\includegraphics[scale=\PGscale]{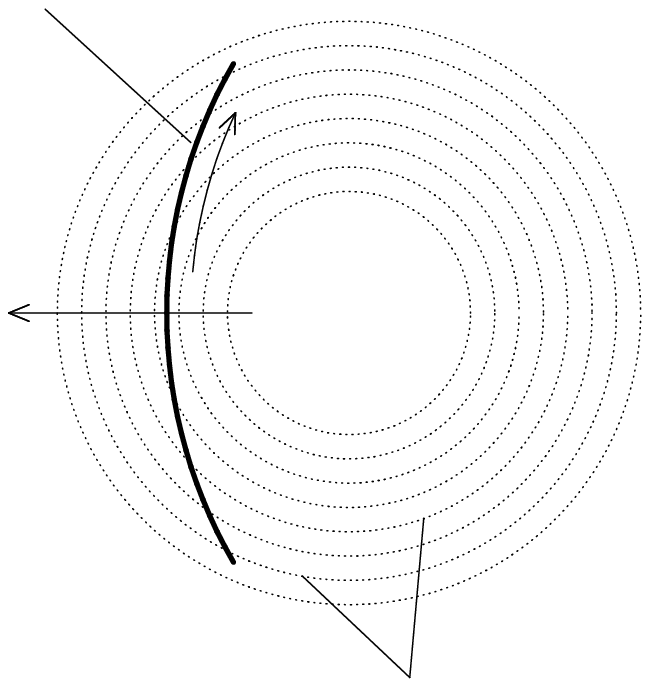}}%
\putlabel{b }{   0.0}{ 550}{3075}{\front}%
\putlabel{t }{   0.0}{2050}{ 275}{\equipotential}%
\putlabel{b }{   0.0}{ 400}{1850}{$\vec{v}$}%
\putlabel{  }{  75.0}{1497}{2724}{\pdecrease}%
 \end{pspicture}%

%% file: mag_tex.tex
\setlength{\figwidthdef}{ 7.0000000000in}%
\setlength{\figheightdef}{ 3.5000000000in}%
\setuplengths%
 \begin{pspicture}(3600,3600)%
 \rput(1800,1800){\includegraphics[width=\PGscale\psfigwidth,height=\PGscale\psfigheight]{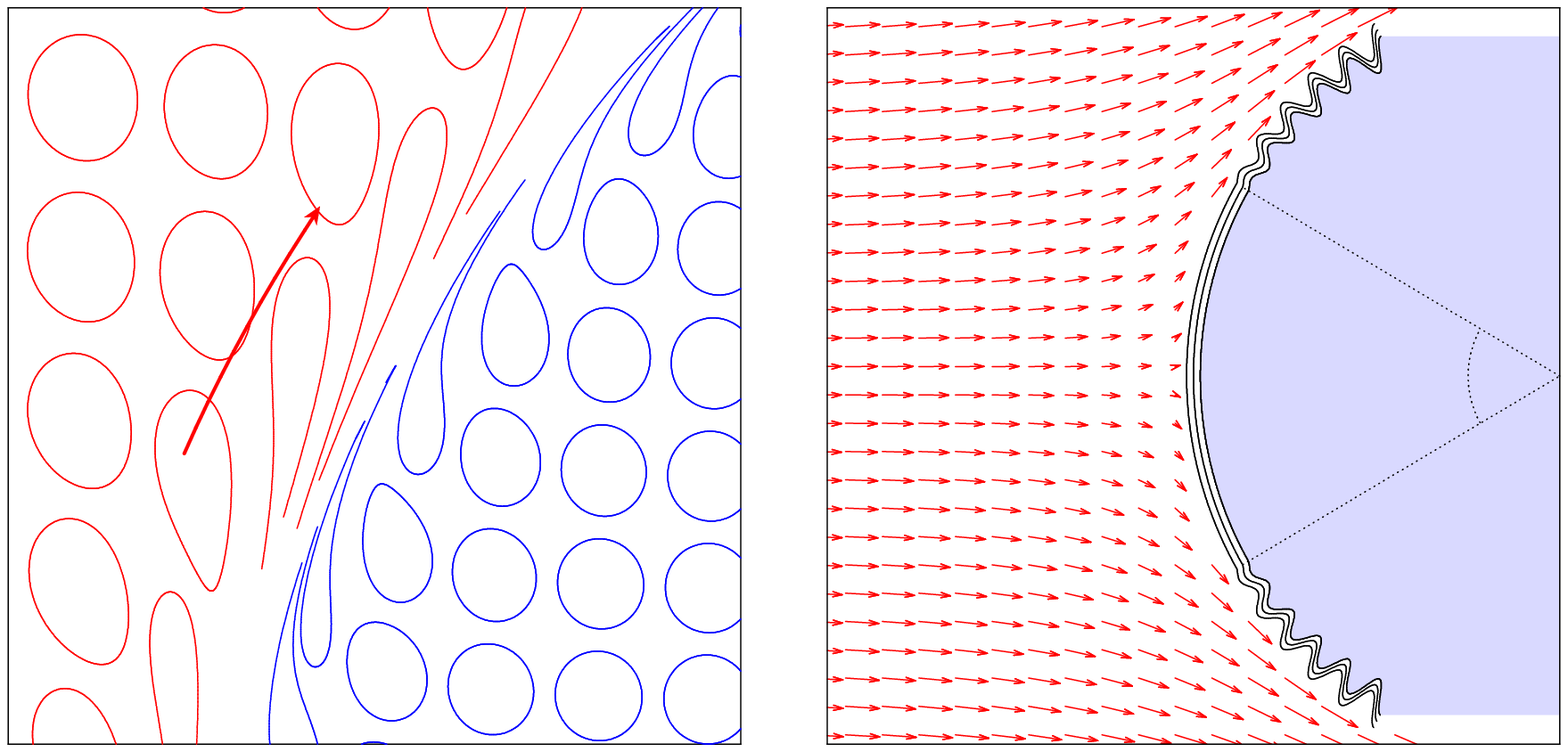}}%
\newrgbcolor{pgcolor}{
   0.  0.  0.}%
\newrgbcolor{pgcolor}{
     1.00000  0.  0.}%
\newrgbcolor{pgcolor}{
   0.  0.    1.00000}%
\newrgbcolor{pgcolor}{
     1.00000  0.  0.}%
\newrgbcolor{pgcolor}{
     1.00000  0.  0.}%
\newrgbcolor{pgcolor}{
   0.  0.  0.}%
\newrgbcolor{pgcolor}{
   0.  0.  0.}%
\newrgbcolor{pgcolor}{
   0.  0.  0.}%
\newrgbcolor{pgcolor}{
   0.  0.  0.}%
\putlabel{  }{  43.0}{2871}{3129}{\vttext}%
\newrgbcolor{pgcolor}{
   0.  0.  0.}%
\putlabel{r }{   0.0}{3357}{1800}{$2\varphi_{\rm cr}$}%
 \end{pspicture}%

%% file: kh-disp-sketch_tex.tex
\setlength{\xunitlen}{0.0013888889in}%
\setlength{\yunitlen}{0.0005555556in}%
\setlength{\runitlen}{0.0008784105in}%
\psset{xunit=\PGscale\xunitlen}%
\psset{yunit=\PGscale\yunitlen}%
\psset{runit=\PGscale\runitlen}%
 \begin{pspicture}(3600,3600)%
 \rput(1800,1800){\includegraphics[scale=\PGscale]{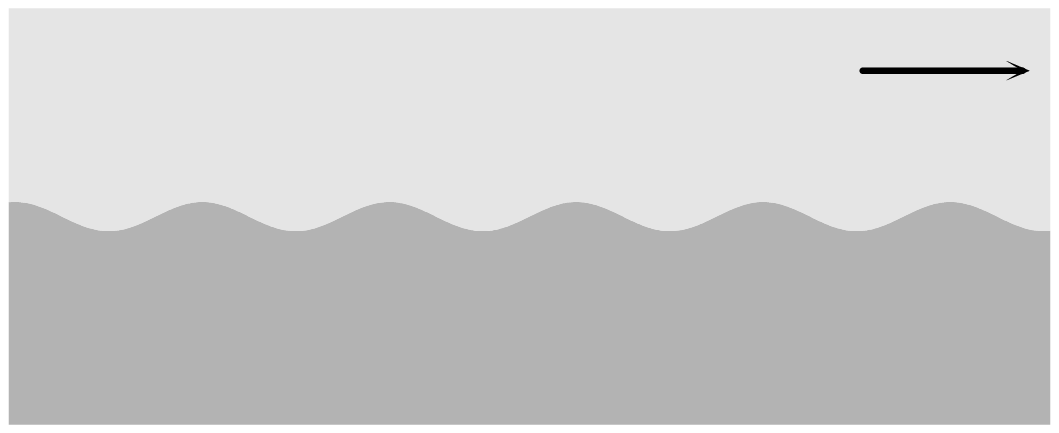}}%
\newrgbcolor{pgcolor}{
    0.701961   0.701961   0.701961}%
\newrgbcolor{pgcolor}{
    0.898039   0.898039   0.898039}%
\newrgbcolor{pgcolor}{
   0.  0.  0.}%
\putlabel{lb}{   0.0}{ 420}{2550}{\Hot}%
\putlabel{lb}{   0.0}{ 420}{1050}{\Cold}%
\putlabel{lt}{   0.0}{ 479}{2475}{\Hotgas}%
\putlabel{lt}{   0.0}{ 479}{ 960}{\Coldgas}%
\putlabel{b }{   0.0}{3000}{2955}{$\vec{v}$}%
\putlabel{t }{   0.0}{3000}{2745}{$M=v/c_h$}%
\putlabel{b }{   0.0}{1800}{1874}{$e^{i(kx-\omega t)}$}%
 \end{pspicture}%